\documentstyle[preprint,aps,epsf,floats]{revtex}
\tighten
\input {amssym.def}
\input {amssym.tex}
\everymath{\rm}
\everydisplay{\rm}

\begin{document}
\draft

\title{Higgs Boson in Superconductors}

\author{C.M.Varma}
\address{Bell laboratories, Lucent Technologies, \\
Murray Hill,NJ 07974}

\maketitle
\begin{abstract}

Superfluid helium, describable by a two-component order parameter, exhibits only
the Bogolubov mode with energy $\rightarrow 0$ at long wavelengths,
while a Lorentz-invariant
theory with a two-component order parameter
 exhibits a finite energy mode at long wavelengths (the Higgs Boson), besides the
above mass-less mode. The mass-less mode moves to high energies if it couples
to electromagnetic fields (the Anderson-Higgs mechanism). Superconductors, on the
 other hand have been theoretically and experimentally  shown to exhibit both modes.
This occurs because the excitations in superconductors have an (approximate)
particle-hole symmetry and therefore show a similarity to Lorentz-invariant theories.

\end{abstract}

\subsection{Introduction}
 
The order parameter for superfluidity or superconductivity  may be written as
\begin{equation}
\psi({\bf k}) = | \psi ({\bf k})| \exp(i \phi ).
\end{equation}
The Free energy is invariant to the value of $\phi$. Then in a neutral superfluid,
 the deviation of the phase $\delta \phi (\bf {k},\omega)$ from  a chosen value
 $\phi = \phi_0$ follows a wave equation with the dispersion $\omega \propto k$
 at long wavelengths. These facts were realized by Bogolubov,\cite{bo}, Anderson \cite{anderson},
  Nambu \cite{nambu}  and others
 and are in accord with the more general Goldstone theorem for collective fluctuations
 in models with continuous broken symmetry.
 
 It was also realised by Anderson that in a charged superconductor the phase mode moves
 to the plasma frequency of the metal, since the current $\bf{j}$,  proportional
 to $\nabla \phi$ is related to the charge density fluctuations through the
 continuity equation. This is also the Higgs mechanism for mass generation in unified
 electro-weak field theories. Since in condensed matter physics, experiments are abundant,
 the underlying physics is often understood more vividly. Indeed the shift of the phase
 mode to the plasma frequency would appear inescapable; otherwise the color of a
 metal would change below the superconducting transition.
 
 In $U(1)$ field theories and generalization to higher groups Higgs also predicted
 the existence of another massive mode - the so called Higgs Boson. The experimental
 search for the Higgs Boson in particle physics has assumed much importance.
 The equivalent of the Higgs Boson is not found in superfluid He$^4$ but about
 twenty years ago, spurred by some anomalous experimental results \cite{sk}, the theory of the
 equivalent of the Higgs Boson in superconductors and its experimental
 identification was provided \cite{lv},\cite{nambuhiggs}.
 In this note I discuss first why such a mode is not present in superfluid$^4$, will not
 be found in the recently discovered weakly interacting Bose gases but is found in
 superconductors.
 
 \subsection{No Higgs in Superfluid Helium}
 
 For weakly interacting Bosons (and as far as symmetry-related properties
 are concerned, He$^4$, as well)the minimal Hamiltonian for the complex
scalar field $\psi$
 is
 \begin{equation}
 H= \int d {\bf r} \left( | \nabla \psi |^2 +r | \psi |^2 +u | \psi |^4 \right).
 \end{equation}
 For $r<0$, a superfluid condensation leads to a finite thermal expectaion value
 \begin{equation}
 < \psi > = [-r/2u]^{1/2} \equiv \rho_0^{1/2}
 \end{equation}
 where $\rho_0$ is the superfluid density. The potential has the form of an
 inverted Mexican hat as shown in fig.( 1 ) in the space of the real
 and imaginary part of $\psi$. One then expands $\psi$ about the
 condensed value
 \begin{equation}
 \psi ( \bf{r},t) = [ \rho_0 + \delta \rho({\bf r},t) ]^{1/2} \exp (i \phi({\bf r},t) )
 \end{equation}
and uses the non-relativistic Schrodinger equation
\begin{equation}
- i\partial \psi / \partial t = H \psi.
\end{equation}
to find after linearising that
\begin{eqnarray}
\partial \phi / \partial t &=& (\nabla^2 \delta \rho + V \delta \rho)/2\rho_0 \\
1/2\rho_0\partial \delta \rho / \partial t &=& \nabla^2 \phi
\end{eqnarray}
 The latter is just the continuity equation. Solving Eqs. (6,7) to linear order leads
 to a pair of degenerate equations for $\phi$ and $\delta \rho$.
 \begin{equation}
 \frac{\partial^2 ( \phi , \delta \rho/\rho_0 )}{\partial t^2}= \nabla^2
 (\phi ,\delta \rho / \rho_0) + \nabla^4 ( \phi , \rho / \rho_0 ),
 \end{equation}
 which leads to the Bogolubov dispersion
 \begin{equation}
 \omega^2 = k^2 + k^4
 \end{equation}
 There is no other collective mode; the mode for amplitude $\rho$ and phase $\phi$
 have identical dispersion. It is not sufficient to have the
 Mexican Hat potential of Fig. (1) to get
 two collective modes, one for the real and the other for the imaginary
  part of the order parameter.
 
 \subsection{Collective Modes in Lorentz-Invariant Theory}
 
 Contrast the above with the case that the problem is Lorentz invariant, so that the
 space and time derivatives have to be of the same order. More typically, one works with a
 Lorentz invariant Lagrangian
 \begin{equation}
 {\cal L}= | \partial_\mu \psi |^2 + r | \psi |^2 + u | \psi |^4 ,  \mu=({\bf r},t).
 \end{equation}
 This has of-course the same potential, of the form of a Mexican hat for $r<0$ as
 in fig. (1). But in contrast to the Lorentz-Invariant theory, besides the
 Goldstone mode which is massless at long wavelengths(just as the Bogolubov mode
 in the non-relativistic theory), there also exists a massive mode for deviation of
 amplitude of $|\psi |$ about the stationary value $ (-r/2u)^{1/2}$ whose
 energy at long-wavelength is
 \begin{equation}
 \omega_{Higgs} = u |2\psi_0|^2,
 \end{equation}
 i.e. given by the curvature of the Mexican hat potential of fig. (1).
 If $\psi$ couples to the electromagnetic field, the phase mode moves up to high energy,
 alleviating a major headache of particle-theorists in the 1960's.
 In SU(2) or higher group
 gauge symmetric theories, the basic physics is the same; depending
 on the structure of the theory
  some "phase" modes may remain massless while there may be a multitude of
  amplitude or Higgs modes. The amplitude or Higgs mode is yet to be discovered
in elementary particle physics.

  We see that with the same effective potential, the Schrodinger equation and the Lorentz-
  invairiant theory give quite different answers. Why should there then be an
  amplitude mode in superconductors as in the Lorentz-invariant theory?
  We will come back to this question after reviewing briefly the microscopic
  derivation of the Higgs in superconductors.
   \subsection{Microscopic theory of the Amplitude Mode in Superconductors}

   The theory of the phase mode in superconductors was developed soon after the
   BCS theory in order to show the gauge invariance of the theory.
No attention was paid to
   the possibility of the amplitude mode, presumably because the experiments did not call
   for one. In 1980 Raman scattering experiments in the compound $NbSe_2$ appeared.
   $NbSe_2$ has a charge density wave (CDW)transition
    at 40 K and a superconducting transition at
   7.2 K. Because of the doubling of the unit cell due to the change in periodicity, a
   new set of vibrational modes appear
   at long wavelengths and finite energie below the
   CDW tranition, see fig. (2). Some of these modes do not
    carry any dipole moment and are therefore  seen in Raman scattering experiments but not
    in optical absorption experiments. It was found that below the superconducting transition
    the spectral weight of such modes
    is partially transferred
     to a sharp mode mode  at an energy of about twice the superconducting gap
      $2 \Delta (T)$. This is shown in fig. (3). Not only is the sum of the spectral
      weight in the two modes conserved, the ratio of the weight in the two peaks could be
      altered by applying a magnetic field which suppresses the superconducitng gap.

      The microscopic theory of this phenomena \cite{lv} was developed
      along the following lines:
      In terms of the two component vectors
\begin{equation}
\Psi_{\bf k} =
\left(
\begin{array}
 c_{{\bf k}\uparrow} \\
\left. c_{{\bf -k}\downarrow}^\dagger \right.
\end{array}
\right)
\Psi_{\bf k}^\dagger =
\left( c_{{\bf k}\uparrow}^\dagger c_{{\bf -k}\downarrow} \right)
\end{equation}
      The hamiltonian of a system of fermions interacting with a non-retarded
      potential is
\begin{eqnarray}
{\cal H} &=& \sum_{\bf k} \Psi^\dagger ({\bf k}) \tau_3 \Psi_{\bf k} \\
&+& \sum_{\bf k,k'.q} V( \bf{k,k',q}) \Psi^\dagger_{{\bf k+q}} \tau_3 \Psi^\dagger_{\bf k}
\Psi^\dagger_{\bf k'-q} \tau_3 \Psi^\dagger_{\bf k'},
\end{eqnarray}
    where the $\tau $'s are Pauli matrices in the space of (12).
    We may write
\begin{eqnarray}
{\cal H} &=& {\cal H}_{BCS} + {\cal H}_1 ,\\
{\cal H}_{BCS} &=& \sum_{\bf k} \Psi^\dagger_{\bf k} \left( \epsilon_{\bf k} \tau_3 + \Delta_{\bf k}
\tau_1 \right)\Psi_{\bf k}.
\end{eqnarray}
    In (16), the amplitude of the gap parameter has been arbitrarily chosen in the $\tau_1$
    direction and phase in the $\tau_2$ direction. Identical physical results
should be obtained for any choice of phase in the
    $\tau_1-\tau_2$ plane. Long wavelength variations about this condition
\begin{equation}
\Psi \rightarrow \exp(i \alpha ({\bf r},t) \tau_3) \psi
\end{equation}
should satisfy the continuity equation for slowly vrying $\alpha ( {\bf r},t)$:
\begin{equation}
\frac{\partial }{\partial t}(\psi^\dagger \tau_3 \psi) + \nabla . \Psi^\dagger
\frac{\bf{P}}{m} \Psi =0.
\end{equation}
This guarantees the existence of the Anderson-Bogolubov mode. But since the
longitudinal part of the electromagnetic field couples to the charge density
$\Psi^\dagger \tau_3 \Psi$, this mode is pushed to the plasma frequency.

The Hamiltonian $\cal H$ obeys another invariance relation besides Eq. (17)\cite{nambu,lv}.
It is invriant to the non-unitary transformation:
\begin{eqnarray}
\Psi & \rightarrow & \exp (\alpha ({\bf r},t)\tau_1) \Psi \\
\nabla & \rightarrow & \nabla + \alpha ( {\bf r},t) \tau_1,
\end{eqnarray}
leading to the pseudo-continuity equation
\begin{equation}
i \Psi^\dagger \tau_1 \left( \frac{\stackrel{\leftarrow}{\partial}}{\partial t} -
\frac{\vec{\partial}}{\partial t} \right) \Psi \\
+ \nabla . \Psi^\dagger \tau_2
\left( \frac{\stackrel{\leftarrow}{P}}{m} + \frac{\vec{P}}{m} \right) \Psi =0.
\end{equation}
A treatment of ${\cal H} - {\cal H}_{BCS}$
  obeying this continuity equation
  reveals the existence of the amplitude mode
  (so-called because its eigenvector is proportional
  to $\tau_1$),
  i.e. the Higgs Boson. At $q=0$, its energy $\nu$ is given by the solution of
  \begin{equation}
  1 + V \sum_{\bf k}
\frac{\epsilon_k^2}{E_k( \nu^2 /4 - E_ k^2)} =0,
  \end{equation}
where $V$ is the appropriate partial wave component of the potential.

For s-wave superconductors with $\Delta$ independent of $k$, the solution is
$\nu =2 \Delta$. For $q << k_F$, the mode is found at
  \begin{equation}
  \nu_q^2 \approx 4 \Delta^2 + v_F^2 q +( \pi^2 /12) i \Delta v_Fq.
  \end{equation}

  For d-wave superconductors, or generally when a continuum
  of single-particle excitations exist
  at low energies , the mode is below twice the maximum
  superconducting gap and is heavily overdamped. One might also wish to worry
about the damping of the Higgs
  Boson  in elementary particle physics.

  Since the amplitude mode is a fluctuation of the Cooper pair density (it exist in the $\tau_1$
  channel), there is no coupling to electromagnetic waves.
  The lattice vibrational modes of the the CDW oscillate the density of states at the Fermi-energy
  and therby couple to the superconducting gap and the amplitude mode. This coupling also pushes
   the amplitude mode below $2 \Delta$ and provide it spectral weight. Given the change of the
   CDW and the superconducting transition with pressure, the coupling of the CDW mode and the
   amplitude mode can be found and the observed phenomena has been explained quantitatively
   without any free parameters \cite{lv}.

   I now return to the earlier question of why such a Higgs mode,a property
    of a Lorentz-invariant theory is found in non-relativistic condensed matter physics.

    \subsection{Approximate particle-Hole Symmetry in Superconductors}

    The mathematical reason for the difference in the
    results for the non-relativistic and the Lorentz-invariant theory is of-course
    that in the former the equation of motion is first order in time and in the second,
     it is second-order in time. The reason the equation of motion for superconductivity
     is effectively second-order in time is the particle-hole symmetry of the theory.
     This is obvious from the BCS Hamiltonian, which is formally identical to
      the Dirac Hamiltonian. The elaborations on the BCS theory discussed above preserve
      this invariance.
      Even in the absence of Lorentz-invariance, the approximate
       particle-hole symmetry of the theory requires the
      equation of motion to be second-order.
      (Relaxing particle-hole symmetry in the
      microscopic theory sketched above
      produces a corresponding damping in the dispersion of the amplitude mode.)
       In fact phenemenological
      Lagrangians for superconductors at low tempertures with a second-derivative in time
      have been derived \cite{abraham}. Near $T_c$,
the damping of the modes is usually handled with
      a phenemenological first order time derivative in a Landau-Ginzburg theory,
      making the theory of superconductors and superfluids look alike.
      This is somewhat misleading.

      Just like in the Dirac equation, the theory of superconductivity deals with conserved
      charge as well as unconserved particle number. Correspondingly,
  there is an important physical difference of weakly interacting Bosons or superfluid
  $He^4$ from superconductors. In the latter, the physical variables, density and current are
  constrained by the continuity equation and no other degree of freedom exists (as long as
  we work at energies small compared to the dissociation of the Boson into constituent
  Fermions). In superconductors, there is an additional degree of freedom,
   the Cooper pair density and therefore the possibility of an additional collective mode.

   \subsection{Acknowledgements}

   I am pleased to be contributing this little note to the joyous occasion celebrating Peter
   Woelfle's sixtieth birthday and to his many contributions to Science,
    including work related to that described here:
    collective modes in superfluid $He^3$ \cite{woelfle}. Over the
     years I have had numerous profitable discussions with him in very many problems ranging
      from the collective mdoes discussed here to transport in heavy Fermions to the theory of
      high temperature superconductivity. I have found him a valuable colleague and a
      scientist upholding the highest standards of science.

      Discussions with Chetan Nayak contributed to my decision to write this note.

\subsection{Figure Captions}

Fig. (1): Plot of the potential in Eqs. (2) and (10) as a function of Real and Imaginary
parts of $\psi$.

Fig.(2):  Raman scattering intensity in $NbSe_2$ at $T=9K$, below the
 CDW transition at $40 K$ and above the superconducting transition at $7.2 K$ in
two different symmetries. The
peaks below $100 cm^{-1}$ arise only below the CDW transition. Data from Ref.(\cite{sk}).

Fig. (3): Same as Fig. (2) at $T=2 K$. The intensity of the new peaks grows below the
superconducting transition and at low temperatures their energy is  at approximately
twice the superconducitng gap. Spectral wieght can be transferred back to the mother peaks
by applying a magnetic field.
\end{document}